\documentclass{aa}  

\usepackage{graphicx}
\usepackage{hyperref}
\usepackage{stfloats}
\usepackage{txfonts, xspace}
\usepackage{array}
\usepackage{colortbl}
\usepackage[table]{xcolor}
\usepackage{enumitem} 
\begin{document}

\titlerunning{Photometric stellar masses in Legacy Surveys}
   \title{Photometric stellar masses for galaxies in DESI Legacy Imaging Surveys}

   \author{Ivana Ebrov\'{a}
      	\inst{1}\thanks{\email{ebrova.ivana@gmail.com}}
      	\and
      	Michal B\'{i}lek \inst{2,3}
      	\and
      	Ji\v{r}\'{i} Eli\'{a}\v{s}ek\inst{1}
      	}

   \institute{FZU – Institute of Physics of the Czech Academy of Sciences, Na Slovance 1999/2, Prague 182 21, Czech Republic %\\
%          	\email{}
    	\and
    	Astronomical Observatory of Belgrade, Volgina 7, 11060 Belgrade, Serbia
     	\and
         	Observatoire de Paris, LERMA, Coll\`ege de France, CNRS, PSL University, Sorbonne University, F-75014, Paris\\  	 
         	}

   \date{Received September 15, 1996; accepted March 16, 1997}

 \abstract{
In many areas of extragalactic astrophysics, we need to convert the luminosity of a galaxy into its stellar mass.
In this work, we aim to find a simple and effective formula to estimate the stellar mass from the images of galaxies delivered by the currently popular DESI Legacy Imaging Surveys.
This survey provides an unsurpassed combination of a deep imaging with an extensive sky coverage in up to four photometric bands.
We calibrated the sought formula against a sample of local galaxies observed by the Spitzer Survey of Stellar Structure in Galaxies (S$^4$G) that was directly dedicated to measure the stellar masses.
For the absolute magnitudes $M_g$ and $M_r$ of a galaxy in the Legacy Surveys $g$ and $r$ bands, we find that the stellar masses can be estimated as $0.673M_g - 1.108M_r + 0.996$ with the scatter of 25\%.
Employing more complex functions does not improve the estimate appreciably, even after including the galaxy ellipticity, S\'ersic index, or the magnitudes in different Legacy Surveys bands.
Generally, measurements in $r$ band were the most helpful ones, while adding $z$-band measurements did not improve the mass estimate much. 
We provide a Python-based script \texttt{photomass\_ls.py} to automatically download images of any galaxy from the Legacy Surveys database, create image masks, generate GALFIT input files with well-assessed initial values, perform the GALFIT photometry, and calculate the stellar mass estimate.
Additionally, we tuned another version of the formula to the magnitudes provided by the Siena Galaxy Atlas 2020 (SGA-2020) with a scatter of 29\%.
For both\,--\,our default and SGA-2020 formula, we offer two alternatives derived from different calibrations of S$^4$G masses that were based on different methods and assumptions. 
}

   \maketitle
%

%%%%%%%%%%%%%%%%%%%%%%%%%%%%%%%%%%%%%%%%%%%%
%%%%%%%%%%%%%%%%%%%%%%%%%%%%%%%%%%%%%%%%%%%%
\section{Introduction}

The total stellar mass of a galaxy is a property of essential importance. 
It allows us to estimate the dark matter content of galaxies through the stellar-to-halo mass relation \citep[e.g.,][]{yang03,am06,moster10,behroozi10,wechsler18,girelli20} or directly the strength of the gravitational field through the radial acceleration relation \citep{mcg16,lelli17,li18rsr}.  
In turn, it has a strong influence on all dynamical phenomena inside the galaxy. 
The success of processes such as supernova winds and outflows from active galactic nuclei to remove gas from galaxies depends on the strength of the gravitational field of the galaxy \citep[e.g.,][]{dekel86,maclow99,stringer12,muratov15,wisdom23,sami24}. 
This dependence is presumably the cause of the observed mass-metallicity relation \citep{ma16}. 
The total mass of the galaxy is believed to determine which process is dominant in removing baryonic mass from the host galaxy: winds of active galactic nuclei in the most massive galaxies or supernova explosions for the less massive hosts \citep{dekel2006,heck23}.
The present-day stellar mass is also one of the main quantities determining the current star formation rate \citep{peng10,speagle14,popesso23,looser24}.
Determining the gravitational field of a galaxy is a necessity for the accurate modeling of the evolution of tidal features -- remnants of galaxy mergers -- that helps us to investigate the merging histories of the host galaxies \citep[e.g.,][]{draper12,george17,oh19,mu19,e20sg,probs,bil22}. 
The mass of a galaxy also determines how the galaxy interacts with its environment. 
Massive galaxies, i.e. galaxies with stronger gravitational fields, have stronger influence on motion of their neighboring galaxies and on the intergalactic gas clouds.

Stellar mass, through the surface density, is believed to have a strong influence on the stability and secular evolution of disk galaxies \citep{toomore64,goldreich65,nipoti24,rg24}.
Surface density is a quantity appearing in the Toomre criterion for the stability of a disk. 
A disk which is not stable will undergo secular evolution and form a bar. Bars are suspected of generating spiral arms \citep{sanders76,sellwood88,athanassoula13,diazgarcia19}. 
It has been recognized early on that with Newtonian gravity, disk galaxies require dark matter halos, otherwise the disks quickly become unstable and transform into spheroids \citep{ostriker73}. 
It is then the fraction of baryonic, and mostly the stellar mass, to the dark matter mass in the central parts which determines the stability of a galaxy. 
The speed at which a bar rotates, and thus perturbs the stellar orbits in the disk periodically, provoking the secular evolution, is influenced, among others, also by the dynamical friction on the dark matter halo. 
It has been found that the induced deceleration of the bar rotation speed depends on the central baryonic fraction of the galaxy \citep{fragkoudi21}.

While the stellar mass of a galaxy is an important quantity, accurately estimating it is not a straightforward task.
Various methods have been developed, each with its advantages and limitations.
One class of methods involves laborious dynamical modeling combined with mass decomposition into dark matter, gas, and stellar components.
This approach is challenging due to the significant and often dominant presence of dark matter, which can lead to substantial uncertainties.
This includes the analysis of rotation curves or Jeans or Schwarzschild modeling \citep{jeans1922,schw79,cap08}. 
These methods require spectroscopic or HI observations, which are generally hard to obtain. 

The analysis of rotation curves particularly of dwarf and low-surface-brightness galaxies showed that the centers of their dark matter halos contain cores instead of the cusps suggested by the dark-matter-only simulations \citep{deblok02,deblok10,oh15}. This has sparked long discussions about the cause of the deviation \citep{delpopolo21,boldrini21}, when solutions such as transformation of dark matter halos  by supernova explosions \citep{governato10,dicintio14}, warm \citep{maccio12,delos23}, fuzzy \citep{deng18,burkert20,banareshernandez23}, and self-interacting \citep{vogelsberger12,rocha13,peter13} dark matter, and modifications of the law of gravity have been proposed.

Dynamical models of stellar kinematics, particularly when combined with strong lensing observations, indicate that stellar populations in massive galaxies show that their initial mass function cannot be universal and instead show radially variations \citep[e.g.,][]{treu10,vandokkum10,sonnenfeld18,liang25}. Rotation curves of disk galaxies give an upper limit on the stellar mass of the disk. For high-surface brightness galaxies, this can give useful results on the models of stellar populations since the contribution of dark matter to the dynamics of the inner parts of these galaxies seems negligible \citep{starkman18}.

Dynamical models of elliptical and lenticular galaxies have revealed, for example, that many of these galaxies show a substantial systemic rotation \citep{emsellem11}, the stellar orbits typically show a substantial anisotropy with a nearly oblate velocity ellipsoid \citep{cappellari07}, where the details depend on the mass of the galaxy \citep{santucci22}, and different stellar populations being at different orbits in Fornax dwarf spheroidal \citep{kowalczyk22}. The models can also be useful for determining the masses of supermassive black holes \citep{pilawa14,thater19,gultekin24}.

One approach to estimating stellar mass involves using spectral energy distribution fitting, which relies on stellar population modeling.  
It again requires spectra or multi-band photometry, ideally spatially resolved. 
The outcome of these methods is influenced by uncertainties in the assumed initial mass function, stellar evolution models, and the star formation history of the galaxy. 
Numerous studies mapped the dependence of stellar mass estimates on the assumptions of the method \citep[see e.g.,][ and references therein]{mitchell13,lee25,kin25}. 
Alternative approach of \citet{eskew2012} and \citet{telford20} minimizes dependence on star-formation histories by utilizing resolved color-magnitude diagrams.
However all these methods are arduous and limited by the availability of high-quality photometric or spectroscopic data for the galaxies.

Another possibility, which we follow here, is to estimate the stellar mass from the luminosity of the galaxy in a few photometric bands. 
The mass-to-light ratio depends on the photometric band and on the age and metallicity distribution of the galaxy. 
It is most advantageous to estimate the stellar mass from the luminosity in infrared bands. 
These bands are ideal for tracing the stellar mass distribution in galaxies as they are not affected much by the dust and by the variations in stellar populations. 
Several studies offer prescriptions to convert infrared fluxes to stellar masses \citep[e.g.,][]{eskew2012,hunt19,jarrett23,duey25}. 

Previously widely used near-infrared Two Micron All Sky Survey \citep[2MASS;][]{2mass} covered basically the whole sky. 
However, it is relatively shallow, such that a substantial fraction of the light of the galaxy remains undetected, causing a systematic error. 
Deeper images are supplied by Wide-field Infrared Survey Explorer \citep[WISE;][]{wise} in an all-sky survey with four mid- to far-infrared bands. 
However its spatial resolution is relatively poor.
The mid-infrared Spitzer Survey of Stellar Structure in Galaxies (S$^4$G) utilizes high-resolution deep images but with limited sky coverage. 
It provides high-quality measurements of stellar masses for 2331 nearby galaxies \citep{sheth2010,s4gcat2011} using 3.6 and 4.5\,$\mu$m bands with the Infrared Array Camera (IRAC) the Spitzer Space Telescope. 
S$^4$G uses calibration of the conversion between infrared flux and stellar mass derived in \citet{eskew2012} from high spatial resolution maps of the Large Magellanic Cloud. 
While these stellar masses are believed to be very precise, they are available for much fewer galaxies than one might wish. 

We are often left with estimating the stellar masses from the photometry in the less advantageous bands. 
For a long time, the most universal way to obtain a stellar mass was to extract the luminosity and color of a galaxy from the Sloan Digital Sky Survey (SDSS) and then to find the most probable mass-to-light ratio using the tables of \citet{bell03}.
These tables are based on galaxy evolution model fits to SDSS $ugriz$ and 2MASS $K$-band fluxes. 
Nowadays, the role of the SDSS as a wide-field imaging survey is being surpassed by DESI Legacy Imaging Surveys \citep[hereafter the Legacy Surveys;][]{ls19}, which is a few magnitudes deeper. 
Legacy Surveys cover around 50\% of the sky and reach surface brightness limit around 29\,mag\,arcsec$^{-2}$ in $g$ and $r$ bands \citep[e.g.,][]{gordon24}.
That means a large step in the image depth of wide-field imaging surveys. 
Legacy Surveys reveal significantly more tidal features around galaxies which calls for their proper modeling in order to utilize the state-of-the-art observational data. 
Thus, it is desirable to have a simple procedure to derive stellar masses for large samples of Legacy Surveys galaxies.

In our present work, we are looking for a simple formula that would allow estimating the stellar masses from 2-D S\'ersic fits to Legacy Surveys images of galaxies. 
We calibrate it against the precise stellar masses of the S$^4$G sample of nearby galaxies. 
Out of many formulas we tested, it turns out that one of the simplest possibilities provides a very good fit, providing stellar mass with a scatter of 25\%. 
This formula relies solely on total magnitudes in the $g$ and $r$ bands. 
In Sect.\,\ref{sec:photo}, we outline the procedure we used to extract the photometry of S$^4$G galaxies from Legacy Surveys images. 
In Sect.\ref{sec:fitting}, we find the most effective formula to compute the photometric stellar masses and compare it with alternative approaches to the problem at hand. 
The described photometric procedure and the formula are compiled into a publicly available Python script for stellar mass estimates \texttt{photomass\_ls.py}, which is introduced in Sect.\,\ref{ssec:mass}.
In Sect.\,\ref{ssec:dis}, we discuss the accuracy of the reference stellar masses and the limitations of our method.

%%%%%%%%%%%%%%%%%%%%%%%%%%%%%%%%%%%%%%%%%%%%
%%%%%%%%%%%%%%%%%%%%%%%%%%%%%%%%%%%%%%%%%%%%
\section{Photometry of Legacy Surveys galaxies}  \label{sec:photo}

Here, we describe the procedure we used to perform photometry on Legacy Surveys images of S$^4$G galaxies with GALFIT \citep{galfit2010}\footnote{\url{https://users.obs.carnegiescience.edu/peng/work/galfit/galfit.html}}.
The process consists of the following steps:

\begin{enumerate}
    \item Download Legacy Surveys images; Sect.\,\ref{ssec:images} 
    \item Create an image mask; Sect.\,\ref{ssec:mask}
    \item Construct an input file for GALFIT; Sect.\,\ref{ssec:file}
    \item Run GALFIT and extract output parameters; Sect.\,\ref{ssec:galfit}
\end{enumerate}

%%%%%%%%%%%%%%%%%%%%%%%%%%
\subsection{Downloading Legacy Surveys images} \label{ssec:images}

FITS images in $g$, $r$, and $z$ bands are downloaded from the Legacy Surveys Data Release 10 (DR10)\footnote{The manual is available at \url{https://www.legacysurvey.org/dr10/description/}} in resolution 1024\,$\times$\,1024 pixels and in a sufficiently large field of view for optimal light profile modeling.
The GALFIT manual recommends the size of the fitted region to be at least 2.5 times the size of the galaxy visible by eye. 
In our work, the angular size of a galaxy was represented by the \texttt{amaj(arcsec)} parameter from \citet{s4gcat2011} catalog -- the semi-major axis at level $\mu$(3.6$\mu$m)\,=\,25.5\,mag(AB)\,arcsec$^{-2}$.

For angularly small galaxies, we preserve the default Legacy Surveys plate scale of 0.262\,arcsec\,pixel$^{-1}$. 
For large galaxies, the image edge was downsized by an integer number $f_{\rm{DS}}$ so that the field of view is close to 5\,$\times$\,5 \texttt{amaj}. 
The photometric zero point for the default plate scale is 22.5\,mag and $22.5 - 2.5$log$_{10}(f_{\rm{DS}}^2)$ for the downsized images. 
The center of the image was determined by the galaxy coordinates using the \texttt{get\_icrs\_coordinates} Astropy routine. 

The following steps are repeated for each of the images in the different photometric bands.

%------------------------------------------------------------------
\begin{figure*}
    \centering
    \includegraphics[width=1\linewidth]{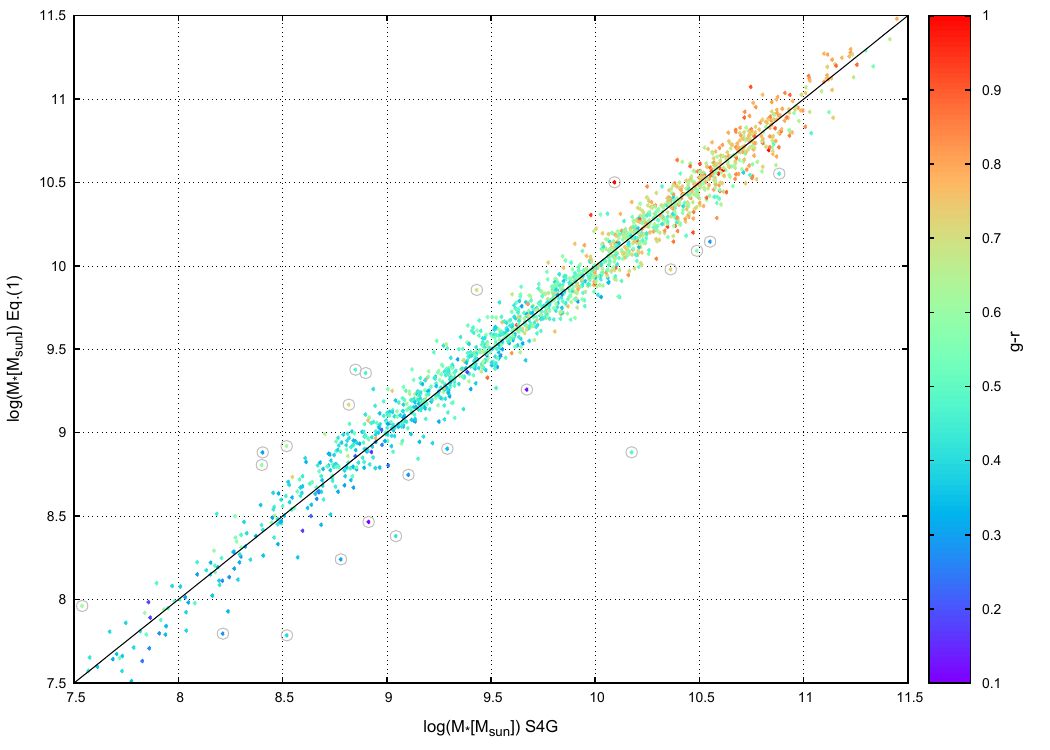}
    \caption{
Correlation of stellar masses drawn from S$^4$G with those computed by Eq.\,(\ref{eq:f}) using magnitudes measured in the Legacy Surveys data. 
The galaxies excluded from the fit, see Sect.\,\ref{ssec:out}, are highlighted by grey circles. 
}
    \label{fig:corr}
\end{figure*}
%------------------------------------------------------------------

%%%%%%%%%%%%%%%%%%%%%%%%%%
\subsection{Masking} \label{ssec:mask}

We create a mask for the image to hide the stars and neighboring galaxies that is subsequently used by GALFIT. 
The final mask is obtained as the union of a coarse and fine mask. 
Both masks are generated by image segmentation using the SEP Python library for Source Extraction and Photometry \citep{sep16,sex}\footnote{\url{https://sep.readthedocs.io}}. 
The SEP algorithm divides the image into individual objects and the background. 
The objects, except for the galaxy of interest, are to be masked.
    
For the coarse mask, the background box sizes are set to be one quarter of the size of the whole image and the object detection threshold is set to one standard deviation of the background noise, in order to mask the faint outskirts of the galaxies. 
The purpose of the coarse mask is to cover the extended objects in the image, typically big galaxies and bright stars. 
    
For the fine mask, the background boxes have the size of two pixels and the detection threshold is set to two sigma. 
The fine mask is supposed to cover small stars and galaxies, which are often missed by the coarse mask. 

It is necessary to exclude the galaxy of interest from each mask. 
Taking advantage of the fact that the galaxy center always aligns with the central pixel of the image, we excluded the object containing this pixel. 
By unifying the two masks, the stars superimposed on the galaxy of interest were also effectively masked. 
However, this approach also masks tiny bright clumps within the galaxies, such as compact star-forming regions or resolved individual stars in nearby dwarf galaxies.

We noted that for galaxies with prominent tidal features, the tidal features were masked as if they were separate objects. 
To prevent this, very elongated objects, namely those whose axial ratio exceeded two, are excluded from the mask. 
The resulting unified mask is saved into a FITS file.

%%%%%%%%%%%%%%%%%%%%%%%%%%
\subsection{Initial parameter values} \label{ssec:file}

Based on the result of the segmentation for the coarse mask, initial guess for the GALFIT fitting was extracted for the galaxy of interest. 
For each detected object, the SEP code directly provides its magnitude and parameters of an ellipse approximating the object: the position angle and the sizes of major and minor axes, denoted by $a$ and $b$, respectively. 
We used their ratio as the initial guess for the axis ratio requested by GALFIT. 
The value of $\sqrt{ab}$ was used as an initial guess for the effective radius of the galaxy. 
For the S\'ersic index required by GALFIT, we universally used the value of four. 
The area of the image to be fitted by GALFIT was determined as three times the extent of the central object as determined by SEP in the settings to create the coarse mask. 

Another component that was involved in the GALFIT model was the sky background, represented by a first order polynomial. 
The initial guess of the central value of the background was chosen to be the median value of the background of the coarse mask provided by SEP. 
The initial values for the tilt of the background were set to zero.

All the parameters, including the plate scale, photometric zero point, and name of the mask file, were saved as an input file for GALFIT.

%%%%%%%%%%%%%%%%%%%%%%%%%%
\subsection{GALFIT modeling} \label{ssec:galfit}

With the input file and mask from the previous step, GALFIT performs the fit with a 2-D single-component S\'ersic profile and sky background gradient. 
The parameters of the GALFIT model were extracted from the GALFIT output, most importantly the parameter “Integrated magnitude,” which was used to derive stellar mass estimates.

%------------------------------------------------------------------
\begin{table*}
\caption{
Fitted formulas
}
\label{tab:fits1}
\centering
\begin{tabular}{lclcccccl}
\hline\hline
Formula & RMS & \( N_{\text{out}} \) & $a$ & $b$ & $c$ & $d$ & Uncertainties \\
\hline
$a M_g + b$ & 0.1703 & 14 & $-0.499$ & $0.427$ & -- & -- & $\pm 0.003, 0.051$ \\
$a M_r + b$ & 0.1271 & 14 & $-0.476$ & $0.597$ & -- & -- & $\pm 0.002, 0.037$ \\
$a M_z + b$ & 0.1788 & 22 & $-0.415$ & $1.668$ & -- & -- & $\pm 0.002, 0.047$ \\
\rowcolor{lightgray} % Highlight row
$a M_g + b M_r + c$ & 0.0972 & 24 & $0.673$ & $-1.108$ & $0.996$ & -- & $\pm 0.019, 0.018, 0.031$ \\
$a M_g + b M_z + c$ & 0.1505 & 20 & $-0.276$ & $-0.191$ & $0.877$ & -- & $\pm 0.010, 0.008, 0.050$ \\
$a M_r + b M_z + c$ & 0.1236 & 17 & $-0.416$ & $-0.054$ & $0.703$ & -- & $\pm 0.007, 0.006, 0.039$ \\
$a M_g + b M_r + c M_z + d$ & 0.0957 & 25 & $0.657$ & $-1.062$ & $-0.028$ & $1.056$ & $\pm 0.019, 0.019, 0.004, 0.032$ \\
\hline
\end{tabular}
\tablefoot{
Parameters, RMS values, and the number of discarded outlying galaxies ($N_\text{out}$) of different fitting formulas for calculating log($M_*$[M$_{\sun}$]) using galaxy total magnitudes in different bands.
}
\end{table*}
%------------------------------------------------------------------

%%%%%%%%%%%%%%%%%%%%%%%%%%%%%%%%%%%%%%%%%%%%
%%%%%%%%%%%%%%%%%%%%%%%%%%%%%%%%%%%%%%%%%%%%
\section{Finding and testing the fitting formula}  \label{sec:fitting}

%%%%%%%%%%%%%%%%%%%%%%%%%%
\subsection{The sample} \label{ssec:sample}

Our parent sample consists of all 1860 nearby galaxies from S$^4$G \citep{sheth2010,s4gcat2011} that have mean redshift-independent distance listed in the S$^4$G catalog. 
After excluding galaxies with missing or incomplete data in the Legacy Surveys and several additional cases with images severely affected by nearby bright stars or those significantly overlapping with another galaxy, our final sample consists of 1732 galaxies. 
This sample was employed to investigate the relation between absolute magnitudes in $g$, $r$, and $z$ bands and the stellar mass, log($M_*$), drawn from S$^4$G.

Integrated magnitudes were extracted from the Legacy Surveys DR10 by the procedure in the previous section, Sect.\,\ref{sec:photo}, and corrected for the Galactic extinction as described in Sect.\,\ref{ssec:mass}.
Absolute magnitudes, $M_g$, $M_r$, and $M_z$, were computed using the same mean redshift-independent distances that were used in S$^4$G to derive the stellar masses.
The dataset with all related parameters is available at Zenodo\footnote{\url{https://doi.org/10.5281/zenodo.14253992}} and CDS.

%%%%%%%%%%%%%%%%%%%%%%%%%%
\subsection{The most effective formula} \label{ssec:formula}

To determine the parameters of the fitting formula, we employed the \texttt{scipy.optimize.curve\_fit} function from the SciPy library\footnote{\url{https://docs.scipy.org/doc/scipy/reference/generated/scipy.optimize.curve_fit.html}}. Initially, the fit was performed on the entire sample. Outliers beyond 3$\sigma$ -- where $\sigma$ is the standard deviation of the residuals from the fit -- were discarded. 
The fitting process was then repeated on the refined dataset until there were no more additional outliers.
The final parameter values and Root Mean Square (RMS) values of the respective formula are presented in Tab.\,\ref{tab:fits1}. 
The RMS scatter can be expressed in terms of percentages as $10^\text{RMS}-1$.

We tried different combinations of magnitude colors. 
The most effective turned out to be the combination of $g$ and $r$ magnitudes, $a M_g + b M_r + c$ (highlighted in the table), which gives a significantly better fit than other two combinations of two-color combinations or single-color fits. 
Thus, we recommend to use the relation
\begin{equation}
\mathrm{log}(M_*[\mathrm{M}_{\sun}]) = 0.673M_g - 1.108M_r + 0.996.
\label{eq:f}
\end{equation}
Comparison of S$^4$G stellar masses with those recovered by Eq.\,(\ref{eq:f}) are shown in Fig.\,\ref{fig:corr}.

Including all three colors improves the fit only slightly (from the scatter of 25.1\% to 24.6\%) on the expense that the $z$-band images may not be available for some galaxies.
Interestingly enough, using solely $M_r$ is as good as or even better than the two two-color combinations that contain $M_z$. 
Even though, from the three bands, the $z$ is the closest to S$^4$G data, we found the measurements in the $z$ band to be generally the least helpful in deriving the proper stellar masses, see also Sect.\,\ref{ssec:quant}.
We also tried a second-degree polynomial equation with cross terms incorporating all colors (i.e. 10-parameter formula; see Appx.\,\ref{apx:forms}), but that improves the RMS only slightly, to 24.0\%.

%------------------------------------------------------------------
\begin{figure*}
    \centering
    \includegraphics[width=\linewidth]{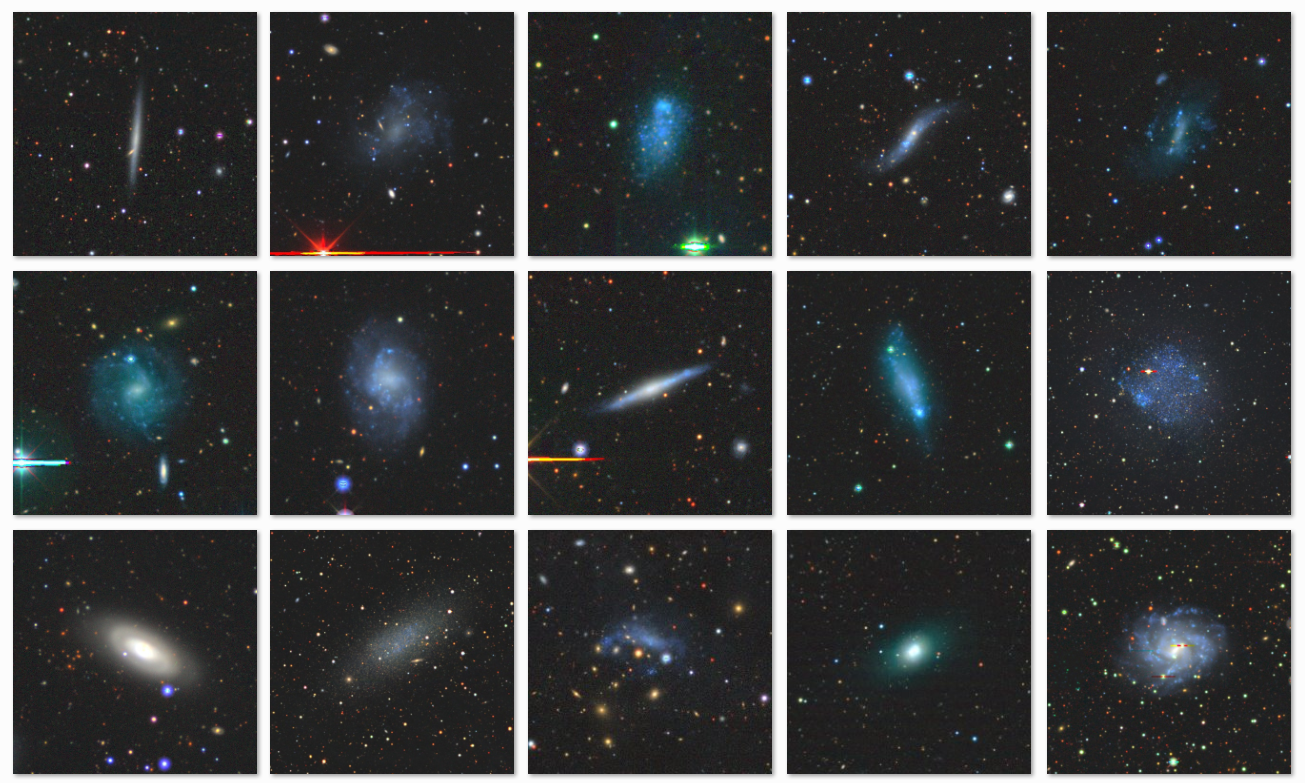}
    \caption{
The 15 most deviant galaxies relative to the relation described by Eq.\,(\ref{eq:f}).
}
    \label{fig:15w}
\end{figure*}
%------------------------------------------------------------------

%------------------------------------------------------------------
\begin{figure*}
    \centering
    \includegraphics[width=\linewidth]{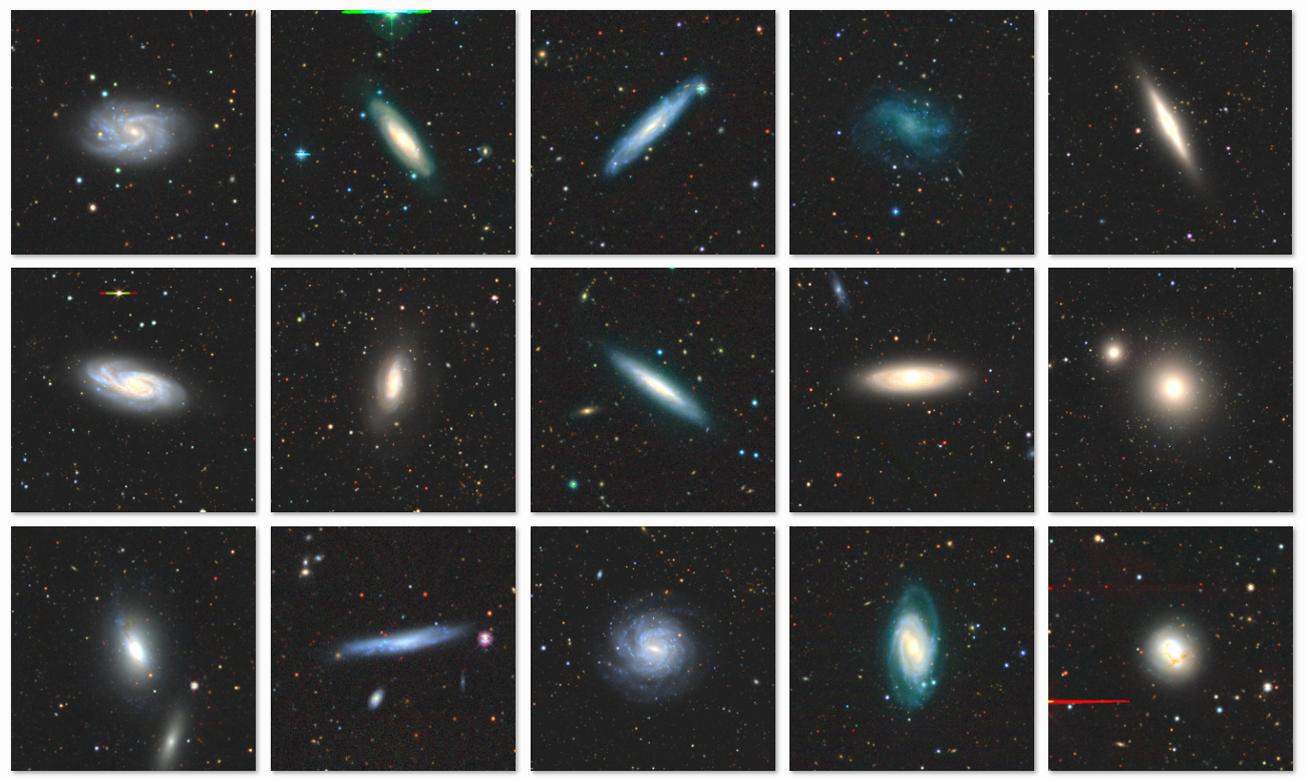}
    \caption{
The 15 galaxies that are most closely aligned with the relation Eq.\,(\ref{eq:f}).
}
    \label{fig:15b}
\end{figure*}
%------------------------------------------------------------------

%%%%%%%%%%%%%%%%%%%%%%%%%%
\subsection{Outlier galaxies} \label{ssec:out}

The number of outlying galaxies for formulas in Tab.\,\ref{tab:fits1} was between 14 and 25; 24 in case of Eq.\,(\ref{eq:f}).
For formulas listed in Tab.\,\ref{tab:fits2}, it was from 23 to 25 galaxies. 
Fig.\,\ref{fig:15w} shows the 15 biggest outliers of Eq.\,(\ref{eq:f}).

Galaxies that were cast aside during searching for the formulas, were often blue faint galaxies, thin faint edge-on galaxies, close dwarf irregulars, and a couple of late-type galaxies with rings (one example can be seen in the bottom left corner of Fig.\,\ref{fig:15w}).

That is not to say that all galaxies of these types have bad estimates of the stellar mass when using Eq.\,(\ref{eq:f}). 
We can find a blue faint galaxy and multiple edge-on galaxies among those that fit the formula best, see Fig.\,\ref{fig:15b}. 
Nevertheless, there seem to be some above-described morphological trends for the outlying galaxies.
We investigated the possibility of improving the formula with parameters that would reflect the shape and type of the galaxy, however the improvement proved to be rather small, see Sect.\,\ref{ssec:quant}.

%%%%%%%%%%%%%%%%%%%%%%%%%%
\subsection{Including other quantities} \label{ssec:quant}

As reported in Sect.\,\ref{ssec:out}, outlying galaxies show some trends in their appearance.
Generally, one can expect the mass-to-light ratio to be different for different morphological types. 
For example, disky galaxies usually have higher dust content than ellipticals.  
We decided to try to improve the formula with parameters that would represent the galaxy morphology.

GALFIT provides the axis ratio, which represents the flattening of the projected shape of the galaxy, and the S\'ersic index, which is a good proxy for the galaxy morphological type.
We tried different formulas combining measurements in different bands with absolute magnitudes, but the improvement was negligible, with the lowest RMS of 0.0930\,dex (i.e. 23.9\% scatter), see Appx.\,\ref{apx:forms}.

Moreover, we explored dependence of mass residues (i.e. estimated logarithmic stellar mass minus the reference logarithmic stellar mass) on various quantities and the method does not seem to work poorly in any particular regime and all trends are consistent with zero, see Appx.\,\ref{apx:err} and Fig.\,\ref{fig:err}. 
However, there are indications that very large galaxies (with radii exceeding 500\,arcsec) may have their stellar masses systematically underestimated (by about 0.2,dex), while galaxies with high S'ersic indices tend to show systematic overestimations.

%------------------------------------------------------------------
\begin{table*}
\caption{
Fitted formulas with alternative calibrations
}
\label{tab:calib}
\centering
\begin{tabular}{lclccccl}
\hline\hline
Formula & RMS & \( N_{\text{out}} \) & $a$ & $b$ & $c$ & Uncertainties \\
\hline
\multicolumn{7}{l}{\textbf{Meidt et al. (2014)}} \\
$a M_g + b M_r + c$ & 0.098 & 20 & $0.582$ & $-1.023$ & $0.992$ & $\pm 0.020, 0.019, 0.030$ \\
$a M_g + b$ & 0.159 & 18 & $-0.498$ & $0.490$ & -- & $\pm 0.002, 0.047$ \\
$a M_r + b$ & 0.119 & 18 & $-0.475$ & $0.664$ & -- & $\pm 0.002, 0.034$ \\
\hline
\multicolumn{7}{l}{\textbf{Querejeta et al. (2015)}} \\
$a M_g + b M_r + c$ & 0.094 & 21 & $0.627$ & $-1.065$ & $1.062$ & $\pm 0.019, 0.018, 0.030$ \\
$a M_g + b$ & 0.161 & 17 & $-0.499$ & $0.504$ & -- & $\pm 0.003, 0.048$ \\
$a M_r + b$ & 0.119 & 18 & $-0.476$ & $0.680$ & -- & $\pm 0.002, 0.034$ \\
\hline
\end{tabular}
\tablefoot{
Parameters, RMS values, and the number of discarded outlying galaxies ($N_\text{out}$) of different fitting formulas using total magnitudes in $g$ and $r$ bands. The reference S$^4$G stellar masses are computed according either \citet{meidt14} or \citet{querejeta15} prescriptions. 
}
\end{table*}
%------------------------------------------------------------------

%%%%%%%%%%%%%%%%%%%%%%%%%%
\subsection{Other S$^4$G mass calibrations} \label{ssec:calib}

In addition to the S$^4$G stellar masses used as default reference masses throughout this work (derived using the calibration from \citealt{eskew2012}), we also applied our fitting procedure against two alternative sets of S$^4$G stellar masses. 
We used prescriptions from \citet{meidt14} and \citet{querejeta15}. 
Both works proceed from Independent Component Analysis and Chabrier initial mass function when deriving their relation between mass-to-light ratios and Spitzer IRAC [3.6] and [4.5] magnitudes of the galaxies. 

From \citet{meidt14}, we adopted the proposed fixed 3.6\,\textmu m mass-to-light ratio of 0.6.
This single value can be applied simultaneously to old, metal-rich and young, metal-poor populations, with the accuracy as low as about 0.1\,dex.
The second calibration is from \citet{querejeta15}, who proposed a color-dependent 3.6\,\textmu m mass-to-light ratio in their Equation\,(7): $\mathrm{log}(M/L)=-0.339\,([3.6]-[4.5])-0.336$. 
We applied the formulas to [3.6] and [4.5] magnitudes of the S$^4$G sample listed in the \citet{s4gcat2011} catalog.
When converting mass-to-light ratios to stellar masses we adopted sun absolute magnitude of 6.02 in IRAC$_{3.6}$ from \citet{sun}. 
Compared to our default S$^4$G stellar masses (the \citealt{eskew2012} calibration), the stellar masses derived from \citet{meidt14} and \citet{querejeta15} works show median offsets of 0.07 and 0.09\,dex, respectively, towards higher masses.

We recalibrated our photometric fitting formulae using these two alternative sets of stellar masses. The parameters of the best-fitting relations are listed in Table\,\ref{tab:calib}.
The preferred formula fitted to the \citet{meidt14} masses is:
\begin{equation}
\mathrm{log}(M_*[\mathrm{M}_{\sun}]) = 0.582\,M_g - 1.023\,M_r + 0.992,
\label{eq:meidt}
\end{equation}
which has an RMS of 0.098\,dex (i.e. 25.2\% scatter).
For the \citet{querejeta15} calibration, the optimal formula is:
\begin{equation}
\mathrm{log}(M_*[\mathrm{M}_{\sun}]) = 0.627\,M_g - 1.065\,M_r + 1.062,
\label{eq:querejeta}
\end{equation}
with a slightly lower RMS of 0.094\,dex (i.e. 24.2\% scatter).
Naturally, the stellar masses computed by these formulas, from magnitudes measured in Legacy Surveys images, show the same offset as their respective reference stellar masses.

%------------------------------------------------------------------
\begin{table*}
\centering
\caption{
Fitted formulas for SGA-2020 magnitudes
}
\label{tab:sgafits}
\begin{tabular}{lcccccccccccc}
\hline
\hline
Formula & RMS(\( M_{26} \)) & RMS(\( M_{\text{TOT}} \)) & \multicolumn{4}{c}{Parameters (\( M_{26} \))} & \multicolumn{4}{c}{Parameters (\( M_{\text{TOT}} \))} \\
 & & & \( a \) & \( b \) & \( c \) & \( d \) & \( a \) & \( b \) & \( c \) & \( d \) \\
\hline
$a M_g + b$ & 0.1664 & 0.2088 & -0.483 & 0.696 & -- & -- & -0.504 & 0.239 & -- & -- \\
$a M_r + b$ & 0.1317 & 0.1725 & -0.461 & 0.854 & -- & -- & -0.480 & 0.424 & -- & -- \\
$a M_z + b$ & 0.1636 & 0.2126 & -0.411 & 1.713 & -- & -- & -0.415 & 1.587 & -- & -- \\
$\boldsymbol{a M_g + b M_r + c}$ & \textbf{0.1090} & 0.1534 & \textbf{0.719} & \textbf{-1.137} & \textbf{1.293} & -- & 0.643 & -1.083 & 0.841 & -- \\
$a M_g + b M_z + c$ & 0.1479 & 0.1865 & -0.274 & -0.180 & 1.110 & -- & -0.285 & -0.184 & 0.750 & -- \\
$a M_r + b M_z + c$ & 0.1280 & 0.1680 & -0.382 & -0.071 & 0.998 & -- & -0.351 & -0.114 & 0.673 & -- \\
$a M_g + b M_r + c M_z + d$ & 0.1086 & 0.1542 & 0.695 & -1.091 & -0.021 & 1.321 & 0.521 & -0.886 & -0.073 & 0.924 \\
\hline
\end{tabular}
\tablefoot{
Parameters and RMS values of different fitting formulas for calculating log($M_*$[M$_{\sun}$]) using galaxy total magnitudes computed from  \( M_{26} \) and \( M_{\text{TOT}} \) magnitudes in SGA-2020.
}
\end{table*}
%------------------------------------------------------------------

%------------------------------------------------------------------
\begin{figure}
    \centering
    \includegraphics[trim=5 5 20 35, clip, width=\linewidth]{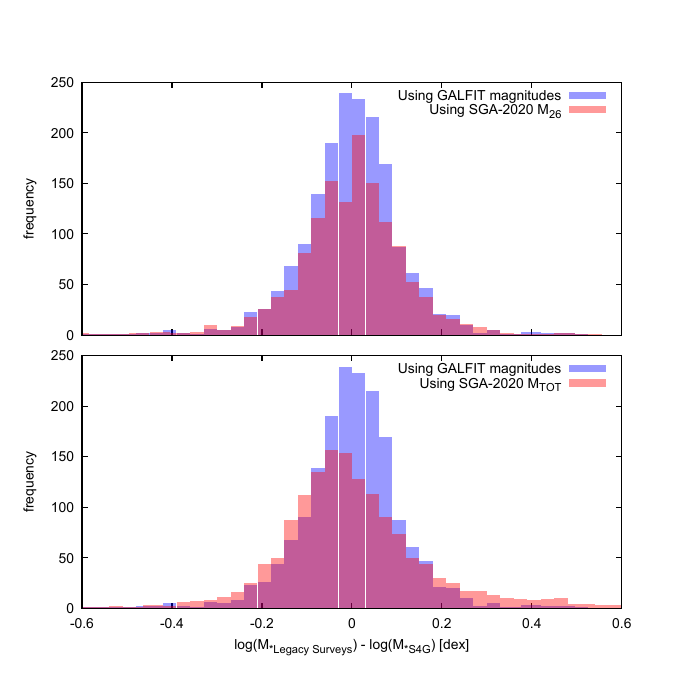}
    \caption{
Histogram of differences between S$^4$G stellar masses and stellar masses recovered using $a M_g + b M_r + c$ formulas: blue histograms\,--\,GALFIT magnitudes with parameters listed in Tab.\,\ref{tab:fits1}, i.e. Eq.\,(\ref{eq:f}); the top red histogram\,--\,using SGA-2020 $M_{26}$, i.e. Eq.\,(\ref{eq:sga}); and the bottom red histogram\,--\,using SGA-2020 $M_\text{TOT}$, parameter values of the formula used with SGA-2020 magnitudes are listed in Tab.\,\ref{tab:sgafits}.
}
    \label{fig:sgahist}
\end{figure}
%------------------------------------------------------------------

%%%%%%%%%%%%%%%%%%%%%%%%%%
\subsection{SGA-2020 magnitudes} \label{ssec:sga}
 
An impressive part of Legacy Surveys data, was already processed and published as the Siena Galaxy Atlas 2020 \citep[SGA-2020;][]{siena} -- a multi-wavelength optical and infrared imaging atlas of 383,620 nearby galaxies constructed from the data of the Legacy Surveys Data Release 9 (DR9).
Although SGA-2020 does not include stellar masses of mass-to-light ratios, it provides measurements of galaxy light profiles in different bands.
We also tried to find suitable formulas using SGA-2020 magnitudes.

We probed two sets of $grz$ magnitudes: \texttt{[G,R,Z]\_COG\_PARAMS\_MTOT} ($M_\text{TOT}$) -- the total magnitude from fit of the curve of growth and \texttt{[G,R,Z]\_MAG\_SB[26]} ($M_{26}$) -- magnitude inside the surface brightness thresholds $\mu_r$\,=\,26\,mag\,arcsec$^{-2}$. 
The parameter $M_\text{TOT}$ was available for 1507 galaxies from our parent sample of 1860 S$^4$G galaxies, $M_{26}$ for 1356. 
In the same way as before, we calculated extinction-correlated absolute magnitudes and found the best fitting parameters for different formulas for the stellar masses.

$M_{26}$ works much better then $M_\text{TOT}$, see Tab.\,\ref{tab:sgafits} and Fig.\,\ref{fig:sgahist}.
The most effective formula was again $a M_g + b M_r + c$ with RMS 0.1090\,dex (0.1534\,dex for $M_\text{TOT}$).
Including the $z$ band brings virtually no improvement, with the RMS of 0.1086\,dex (0.1542\,dex from $M_\text{TOT}$).
Even with the second-degree polynomial (not listed in the table), RMS remains as high as 0.1067\,dex (0.1466\,dex for $M_\text{TOT}$).

Using $M_{26}$ magnitudes recover the S$^4$G stellar masses almost as good as the GALFIT magnitudes (Sect.\,\ref{ssec:formula}), but, as SGA-2020 is drawn from DR9, $M_{26}$ were available only for 73\% of the parent sample, while with DR10 and GALFIT we were able to retrieve magnitudes for 93\%. If in need to use SGA-2020, we recommend using absolute magnitudes computed from \texttt{MAG\_SB[26]} parameters with following formula:
\begin{equation}
\mathrm{log}(M_*[\mathrm{M}_{\sun}]) = 0.719M_g - 1.137M_r + 1.293.
\label{eq:sga}
\end{equation}
Alternatively, formulas for $M_{26}$ calibrated against different S$^4$G stellar masses (see Sect.\,\ref{ssec:calib}) can be used\,--\,with S$^4$G masses according to \citet{meidt14} (RMS 0.1058\,dex):
\begin{equation}
\mathrm{log}(M_*[\mathrm{M}_{\sun}]) = 0.649\,M_g - 1.070\,M_r + 1.327
\label{eq:meidt-sga}
\end{equation}
or according to \citet{querejeta15} (RMS 0.1046\,dex):
\begin{equation}
\mathrm{log}(M_*[\mathrm{M}_{\sun}]) = 0.678\,M_g - 1.097\,M_r + 1.375.
\label{eq:querejeta-sga}
\end{equation}

%------------------------------------------------------------------
\begin{figure}
    \centering
    \includegraphics[width=\linewidth]{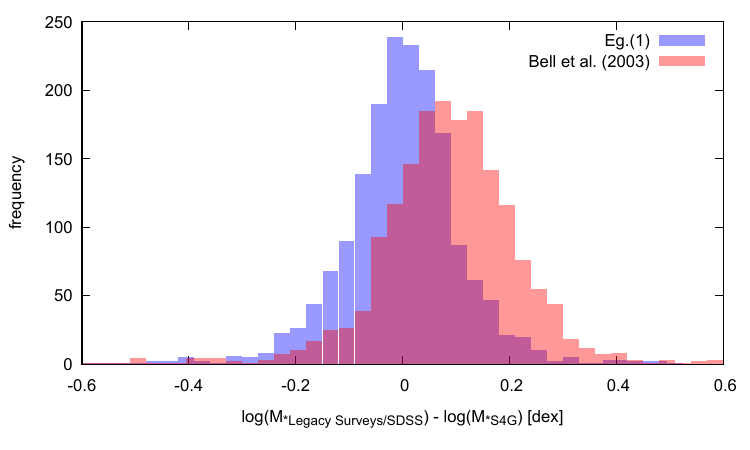}
    \caption{
Histogram of differences between S$^4$G stellar masses and stellar masses computed by Eq.\,(\ref{eq:f}) for the blue histogram and via mass-to-light ratios using tables in \citet{bell03} for the red histogram.
}
    \label{fig:bellhist}
\end{figure}
%------------------------------------------------------------------

%%%%%%%%%%%%%%%%%%%%%%%%%%
\subsection{SDSS formula} \label{ssec:bell}

We compare the performance of our formula, Eq.\,(\ref{eq:f}), with a formula derived by using SDSS and 2MASS in \citet{bell03}.
Specifically, we computed mass-to-light ratios in $g$ band using equation
\begin{equation}
\mathrm{log}(M/L_g) = 0.499 + 1.519\,(g - r)
\label{eq:belleq}
\end{equation}
drawn from Table\,7 in \citet{bell03}.
This equation was designed for SDSS Petrosian magnitudes. 
As the $g$ and $r$ filters at the instruments used by the Legacy Surveys were designed to be similar to SDSS filters, we tried Eq.\,(\ref{eq:belleq}) on the extinction-corrected GALFIT magnitudes of our sample of 1732 galaxies.
The mass-to-light ratios were converted to stellar masses with the absolute magnitudes of the galaxies derived with S$^4$G distances and the absolute magnitude of the Sun in DES\_g = 5.05 drawn from \citet{sun}.

Stellar masses derived in this way reproduce S$^4$G values less precisely than our method.
RMS of the whole sample, except 19 outlier galaxies (outside the three-sigma interval), is 0.1445\,dex (i.e. 39.5\% scatter) when using Eq.\,(\ref{eq:belleq}), while our method gives scatters of 25.1\,--\,25.1\%, using Eq.\,(\ref{eq:f}), Eq.\,(\ref{eq:meidt}), or Eq.\,(\ref{eq:querejeta}).
The systematic median offset between mass estimates from Eq.\,(\ref{eq:belleq}) and our default Eq.\,(\ref{eq:f})\,--\,the \citet{eskew2012} calibration, is 0.09\,dex, see also Fig.\,\ref{fig:bellhist}.
The values are, on average, more consistent with \citet{querejeta15} calibration, Eq.\,(\ref{eq:querejeta}), and a smaller offset (0.02\,dex) against the \citet{meidt14} calibration, Eq.\,(\ref{eq:meidt}). 
However, the scatter of \citet{bell03} tables applied on Legacy Surveys data is still notably higher than those of all three versions of our method.

%%%%%%%%%%%%%%%%%%%%%%%%%%
\subsection{Rest-frame colors} \label{ssec:Kcorr}

The S$^4$G sample concerns nearby galaxies; our subsample of S$^4$G includes galaxies ranging from 0.415 to 94.1\,Mpc, with eight having distances greater than 60\,Mpc.
In previous sections, we neglected wavelength shifts  caused by radial velocities of the galaxies. 
Here, we examine the effect of rest-frame colours. 
First, we tested that excluding these eight most distant galaxies affects the fits negligibly. 
We then proceeded to include $K$ corrections for the entire sample.

As the $g$ and $r$ filters of the Legacy Surveys instruments were designed to be similar to SDSS filters, we employed, for these filters, the $K$-corrections calculator\footnote{\url{http://kcor.sai.msu.ru/}} \citep{kcorr1,kcorr2}, which was designed for SDSS colors. 
Redshifts of galaxies were calculated from heliocentric radial velocities from radio measurement provided by S$^4$G catalog. 
For the forty galaxies in our sample that do not have the radial velocities listed in S$^4$G, we supplied the value from the HyperLeda database\footnote{\url{http://atlas.obs-hp.fr/hyperleda/}} \citep{hyperleda}.

The $K$ correction in the $g$ ($r$) band ranges from -0.0081 (-0.0019) to 0.042 (0.025) for the whole sample and it is smaller than 0.013 (0.012) when excluding the eight galaxies beyond 60\,Mpc. 
We repeated the fitting of the formula $a M_g + b M_r + c$ (see Sect.\,\ref{ssec:formula}) but this time using $K$-corrected magnitudes. 
The fit has the same RMS as Eq.\,(\ref{eq:f}), with only slight changes in the fitted parameter values
\begin{equation}
\mathrm{log}(M_*[\mathrm{M}_{\sun}]) = 0.686M_g - 1.121M_r + 0.994.
\label{eq:Kcorr}
\end{equation}
These changes in parameter values translate to variations in log($M_*$) of at most 0.007\,dex throughout the sample, which this is negligible compared to the scatter in the relation Eq.\,(\ref{eq:f}).

For the purpose of stellar mass estimates of galaxies at non-negligible redshifts, we included an option for $K$ correction, using the $K$-corrections calculator, into the \texttt{photomass\_ls.py} script (see Sect.\,\ref{ssec:mass}). 
Note that the $K$-corrections calculator has the redshift coverage up to 0.5.

%------------------------------------------------------------------
\begin{figure}
    \centering
    \includegraphics[width=\linewidth]{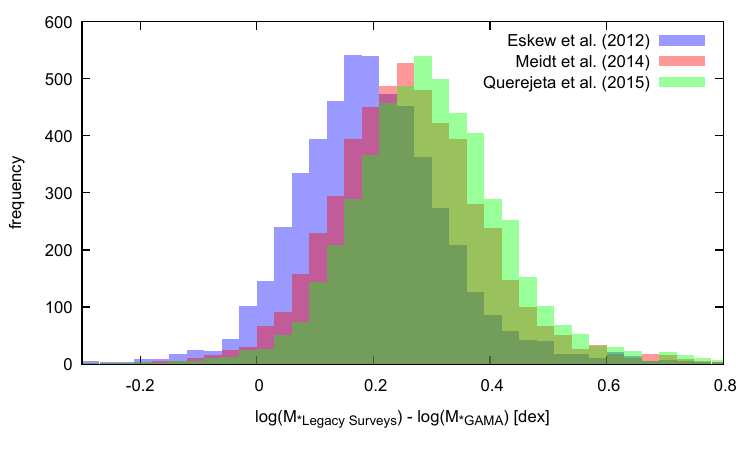}
    \caption{
Histogram of differences between GAMA stellar masses and stellar masses computed from Legacy Surveys images using Eqs.\,(\ref{eq:f}), (\ref{eq:meidt}), and (\ref{eq:querejeta}), which were derived using calibration of S$^4$G stellar masses adopted from \citet{eskew2012}, \citet{meidt14}, and \citet{querejeta15}, respectively. 
Compared to GAMA values, the three sets have offset 0.19, 0.26, and 0.29\,dex, respectively. 
}
    \label{fig:gama}
\end{figure}
%------------------------------------------------------------------

%%%%%%%%%%%%%%%%%%%%%%%%%%
\subsection{Comparison with GAMA} \label{ssec:gama}

We tested our formula on galaxies from the Galaxy And Mass Assembly (GAMA) survey \citep{gama1,gama2}.
Although GAMA has significantly lower spatial resolution than S$^4$G, it includes many more sources with a wider redshift range, making it suitable for testing the impact of the $K$ correction.

We took advantage of the GAMA catalog StellarMassesGKVv24 \citep{gamamass,gamadr4}\footnote{\url{https://www.gama-survey.org/dr4/schema/table.php?id=690}}, which contains stellar mass estimates for 370k sources. 
We limited our sample to sources at redshifts below 0.5, as this is the limit of the K-correction calculator.

To select a suitable subsample of galaxies, we crossmatched the GAMA sources with the Principal Galaxies Catalogue \citep[PGC;][]{pgc} reducing the list to 14.6k items. 
We retained only 11k sources that matched with galaxy-type objects in the NED database and had a redshift difference smaller than 0.0003 between the NED and GAMA catalogs.
We further excluded 3.2k cases where a single NED object matched multiple GAMA sources.

This procedure yielded 7.8k galaxies, which we processed in the same way as S$^4$G galaxies, i.e. using Legacy Surveys DR10 data and the  methodology described in Sects.\,\ref{sec:photo} and \ref{ssec:mass}. 
We used R90 (Approximate elliptical semi-major axis containing 90\% of the flux) from the GAMA gkvScienceCatv02 catalog as the input radius for each galaxy.
Around 2.5k galaxies could not be processed due to missing or temporarily unavailable Legacy Surveys DR10 data, or because GALFIT did not converge.  
However, at this point, we had a sufficient number of stellar mass estimates for a meaningful comparison with the GAMA catalog, and we did not proceed with further processing runs.

We visually inspected the 300 galaxies with the largest discrepancies between the measured $g$ and $r$ magnitudes and discarded about half of them due to incomplete Legacy Surveys data, close overlaps, or images corrupted by nearby bright stars. 
We ultimately obtained reliable stellar mass estimates for 5,187 galaxies using our procedure.

Most of our stellar mass estimates are higher than those listed in the GAMA StellarMassesGKVv24 catalog, with a median offset of 0.19\,dex when applying the $K$ correction, and 0.23\,dex without it. 
As expected, low-redshift galaxies have the same offset in both cases\,--\,0.21\,dex for the 1885 galaxies  with redshifts below 0.07. 
For galaxies at redshifts above 0.17 (282 galaxies), the offset decreases significantly from 0.29\,dex to 0.12\,dex when the $K$ correction is applied.  

Although, in the vast majority of cases, the $g$ and $r$ magnitudes measured in Legacy Surveys data are brighter after applying the correction, the color of the galaxies ($g-r$) is changed in a way that makes the stellar mass estimates mostly lower than those without $K$ correction. 
This effect is the strongest for massive galaxies at higher redshifts.  

All of the above results were obtained using Eq.\,(\ref{eq:f}), which is based on the \citet{eskew2012} calibration of S$^4$G masses.
When applying Eq.\,(\ref{eq:meidt}), derived from \citet{meidt14} calibration, and Eq.\,(\ref{eq:querejeta}) with \citet{querejeta15} calibration (see Sect.\,\ref{ssec:calib}), the offsets are even higher, 0.26 and 0.29\,dex, respectively (both with the $K$ correction). 
Fig.\,\ref{fig:gama} presents the distributions of residuals for all three calibrations (with the $K$ correction applied).

There are only 14 galaxies that appear in both samples that we used\,--\,5.2k GAMA and 1.7k S$^4$G galaxies. 
For these 14 galaxies, we can compare GAMA masses and masses computed directly from 3.6 and 4.5\,$\mu$m bands of the Spitzer data. 
In all three calibration cases, the S$^4$G-based stellar masses show systematic offsets toward higher values, with median differences around 0.3\,dex. 

A possible factor creating these offsets can be the depth and resolutions of the data, which are significantly higher for the S$^4$G survey. 
Interestingly, the offset is higher for \citet{meidt14} and \citet{querejeta15} calibrations that assume the Chabrier initial mass function, which was also used to compute the stellar mass estimates of GAMA galaxies. 

A possible contributor to these offsets is the difference in data quality: S$^4$G provides significantly deeper and higher-resolution imaging than is available in the GAMA survey. 
Interestingly, the offsets are larger for the \citet{meidt14} and \citet{querejeta15} calibrations, despite their use of the Chabrier initial mass function, which was also adopted in the GAMA stellar mass estimates. 
This suggests that differences in methodology and photometric processing, in addition to assumptions of the initial mass function, may play a role in the observed discrepancies.

%%%%%%%%%%%%%%%%%%%%%%%%%%%%%%%%%%%%%%%%%%%%
%%%%%%%%%%%%%%%%%%%%%%%%%%%%%%%%%%%%%%%%%%%%
\section{Script for stellar mass estimates} \label{ssec:mass}

We provide a Python script named \texttt{photomass\_ls.py}, which is publicly available at GitHub\footnote{\url{https://github.com/PidiGalaxies/photomass}}. 

The script automates the galaxy processing steps described in Sect.,\ref{sec:photo} and includes a stellar mass estimation based on the formula, Eq.\,(\ref{eq:f}), derived in Sect.\,\ref{sec:fitting}, using the $g$ and $r$ magnitudes obtained by the script. 
In addition, two alternative estimates are produced using Eqs.\,(\ref{eq:meidt}) and Eq.\,(\ref{eq:querejeta}) derived in see Sect.\,\ref{ssec:calib}.
These two estimates are based on different sets of the reference S$^4$G masses that were based on different methods and assumptions. 

The primary inputs to the script are the galaxy name and its angular size in arcseconds, e.g. the \texttt{amaj} parameter.
Optionally, the user can provide the galaxy distance in Mpc, which is used to compute the stellar mass estimate.
The distance can be accompanied with custom coordinates of the galaxy as an additional input parameter. 

For the given galaxy, the script downloads DR10 Legacy Surveys images in $g$ and $r$ bands (Sect.\,\ref{ssec:images}), creates an image mask (Sect.\,\ref{ssec:mask}), constructs an input file for GALFIT (Sect.\,\ref{ssec:file}), and runs GALFIT and reads its output parameters (Sect.\,\ref{ssec:galfit}). 
The GALFIT magnitudes are corrected for Galactic extinction from \citet{ext11}, in the respective filter, using NASA/IPAC Extragalactic Database (NED)\footnote{\url{https://ned.ipac.caltech.edu}}.
In case the galaxy distance is provided as an input parameter, the absolute magnitude is computed using this distance, otherwise, the redshift downloaded from NED is used. 
Once the absolute magnitudes in $g$ and $r$ bands are determined, the stellar mass of the galaxy is estimated via the formula Eq.\,(\ref{eq:f}), derived in Sect.\,\ref{ssec:formula}. 
Two alternative mass estimates are calculated using Eqs.\,(\ref{eq:meidt}) and Eq.\,(\ref{eq:querejeta}) derived in see Sect.\,\ref{ssec:calib}.

The \texttt{photomass\_ls.py} script provides additional optional functionalities, including photometry in the $z$ and $i$ bands of Legacy Surveys images and $K$ correction for rest-frame colors of galaxies with redshifts up to 0.5.
An example of script usage is provided in Appx.\,\ref{apx:ex}.l 
The full manual, with additional examples, is available at GitHub\footnote{\url{https://github.com/PidiGalaxies/photomass}}.

%%%%%%%%%%%%%%%%%%%%%%%%%%%%%%%%%%%%%%%%%%%%
%%%%%%%%%%%%%%%%%%%%%%%%%%%%%%%%%%%%%%%%%%%%
\section{Discussion} \label{ssec:dis}

%%%%%%%%%%%%%%%%%%%%%%%%%%
\subsection{S$^4$G stellar masses} 

We calibrated our formulas against the reference sample of S$^4$G galaxies.
Stellar masses of galaxies in S$^4$G were derived from fluxes in the mid-infrared bands of the Spitzer Space Telescope.
These bands are well suited for tracing the stellar mass distribution in galaxies, as they are not affected much by the dust and by the variations in stellar populations.
Reported uncertainties of the Spitzer photometry for the S$^4$G sample span from 0.001 to 0.339\,mag, with the vast majority of the sample having the errors smaller than 0.03\,mag.

Our default set of reference stellar masses is based on the calibration of \citet{eskew2012}\,--\,the conversion between the fluxes and stellar masses from high spatial resolution maps of the Large Magellanic Cloud. 
This calibration minimizes dependence on star-formation histories by utilizing resolved color-magnitude diagrams.
Several studies reported a good agreement between this method and stellar mass estimates obtained from other approaches, such as spectral energy distribution fitting; namely \citet{cybulski14} using estimates from \citet{kauffmann03}, \citet{meidt14}, \citet{querejeta15}, or \citet{duey25}. 

\citet{eskew2012} reported the intrinsic uncertainty in their stellar mass estimates is 0.12\,dex ($\sim$30\%) based on the region-to-region scatter throughout the Large Magellanic Cloud. 
This scatter is primarily attributed to the contamination by young stars and hot dust. 
As the random fluctuations average out when summing over larger regions containing more stars, \citet{eskew2012} achieved a precision of a few percent for the total stellar mass of the Large Magellanic Cloud. 
The random fluctuations in contamination are expected to average out to some extent for unresolved galaxies as well, making $\sim$30\% a reasonable estimate of the random uncertainty in stellar mass estimates for S$^4$G galaxies.

Using mid-infrared bands mitigates much of the unwanted effects of dust, and the calibration of \citet{eskew2012} minimizes dependence on star-formation histories.
However, galaxies with extreme metallicities or high specific star-formation rates may deviate from this calibration.

%%%%%%%%%%%%%%%%%%%%%%%%%%
\subsection{Systematic uncertainties of the reference sample} 

Probably the largest systematic uncertainty comes from the choice of the reference sample of stellar masses itself. 
Every method that derives stellar masses or mass-to-light ratios comes with its own set of biases and assumptions. 
The \citet{eskew2012} calibration was derived by studying resolved color-magnitude diagrams of the Large Magellanic Cloud, utilizing Spitzer 3.6 and 4.5\,$\mu$m magnitudes.
That means that the calibration was derived for one specific galaxy, with its particular star formation history and metallicity.
The authors show that the formula works consistently for different star formation regions within the Large Magellanic Cloud, however they did not explore the dependence of their calibration on metallicity, as there is little variation in metallicity within the galaxy. 
It is not certain how well it performs for galaxies with more complex or different stellar populations. 
Our formula inherits any systematic uncertainties present in the Eskew et al. (2012) calibration.

One particular source of uncertainty for our reference sample lies in the choice of the initial mass function (IMF). 
Different IMFs assume different proportions of low- to high-mass stars, leading to systematic shifts in stellar mass calibrations. 
\citet{eskew2012} adopted a Salpeter IMF and reported that lighter versions yield lower mass estimates (by 0.1\,dex for diet-Salpeter and 0.25\,dex for Chabrier IMF). 
They also noted that, within their framework, these lighter IMFs are disfavoured by dynamical mass measurements of the Large Magellanic Cloud. 
However, the broader astrophysical community has not reached a consensus on the most appropriate IMF for all galaxy types or environments.
While the Chabrier IMF has become a common choice in many extragalactic studies due to its consistency with observations of the Milky Way, the Salpeter IMF is still frequently used and often results in higher stellar mass estimates.
There is also growing evidence that the IMF might not be universal, but instead may vary with galaxy mass, environment, or star formation history (e.g., \citealt{hoversten08,gunawardhana11,marks12,cappellari12}; see also \citealt{Smith20} for a review). 

In addition to IMF, stellar mass estimates are sensitive to several other underlying assumptions that affect both the S$^4$G calibrations and our derived relations. 
One key factor is the assumed star formation history of a galaxy, which governs the mix of stellar ages and thus the luminosity--mass relationship. 
Most stellar population synthesis models rely on simplified or parametric star-formation histories, which may not accurately capture complex or bursty formation histories\,--\,particularly in dwarf, irregular, or recently interacting systems. 
\citet{eskew2012} method tries to minimize dependence on star-formation histories by utilizing resolved color-magnitude diagrams instead.
Dust attenuation and its geometry introduce further uncertainty, especially in optical bands where patchy dust can redden light in ways that mimic older stellar populations. 
While the Spitzer mid-infrared bands used by S$^4$G reduce dust effects, they do not eliminate them entirely, and differences in dust geometry may still bias the mass-to-light ratios. 
Uncertainties in stellar population models themselves, such as treatment of thermally pulsing asymptotic giant branch (TP-AGB) stars, also affect derived masses. 
Finally, the mass estimates can be affected by systematic photometric errors, including sky subtraction and aperture effects. 

To explore these uncertainties, we derived stellar mass formulas using two alternative calibrations of the S$^4$G Spitzer data (see Sect.\,\ref{ssec:calib}). 
Compared to our default \citet{eskew2012} calibration, \citet{meidt14} and \citet{querejeta15} have median offsets of 0.07 and 0.09\,dex, respectively, towards higher masses. 
This discrepancy cannot be explained by differences in the assumed IMF: both \citet{meidt14} and \citet{querejeta15} adopt the Chabrier IMF, which should produce lower mass estimates than the Salpeter IMF used by \citet{eskew2012}.
It is likely that the offset stems from differences in methodology.
Whereas \citet{eskew2012} relied on resolved stellar populations, \citet{meidt14} and \citet{querejeta15} used an Independent Component Analysis approach applied to integrated Spitzer photometry. 
Some discrepancy may also emerge from systematic photometric errors if, for example, the subtraction of background gradients is treated differently. 

\citet{bell03} derived their relation for the mass-to-light ratio using stellar population synthesis with SDSS and 2MASS data. 
They assumed the diet-Salpeter initial mass function. 
The average stellar masses we derived using their relation are consistent with those from the \citet{querejeta15} calibration, see Sect.\,\ref{ssec:bell}.  
However, the scatter of the mass estimates for our sample is notably higher for the \citet{bell03} ralation\,--\,40\%, compared to 24\% for \citet{querejeta15} calibration.
This higher scatter can be easily attributed to the fact that relations of \citet{bell03} were derived for SDSS Petrosian magnitudes, while we are applying them to integrated GALFIT magnitudes of Legacy Surveys images. 
This alone demonstrates that, regardless of systematic uncertainties, for the Legacy Surveys data, it is advantageous to use formulas that were actually derived for those data. 

Somewhat surprising is the comparison with GAMA stellar masses, see Sect.\,\ref{ssec:gama}. 
The GAMA masses are about 0.2\,dex lower than all four of the above-mentioned estimates. 
The \citet{eskew2012} calibration shows the smallest offset (0.19\,dex). 
\citet{meidt14} and \citet{querejeta15} calibrations have higher offsets (0.26 and 0.29\,dex, respectively), even though they assumed the same initial mass function (Chabrier) as the GAMA mass estimates. 
A possible factor behind these offsets could be the depth and resolution of the data, which is significantly higher for the S$^4$G images.

Overall, the differences among the three calibrations of S$^4$G stellar masses indicate that the choice of reference sample introduces a systematic uncertainty of at least 0.1\,dex to our method.

%%%%%%%%%%%%%%%%%%%%%%%%%%
\subsection{Limitations of our method} 

First of all, one should always check the quality of Legacy Surveys images.
In some cases, especially (but not only) near the edges of the survey sky coverage, data in certain filters may be missing or incomplete. 
Additionally, bright nearby stars can contaminate galaxy images. 
Among the galaxies excluded from the original sample, there were several that had a close overlap with a galaxy of a similar or bigger size, resulting in a bad GALFIT model. 

To facilitate a quick quality check, the \texttt{photomass\_ls.py} script automatically downloads JPEG images of the processed galaxy in the selected filters. 
While the Legacy Surveys documentation does not specify a strict saturation limit for objects, we advise caution when working with bright galaxies. 
Our sample includes galaxies with apparent total magnitudes ranging from 8.1 to 16.9 in the $g$ band, and we have not encountered any issues related to saturation.

We explored dependence of mass residues (i.e. estimated logarithmic stellar mass minus the reference logarithmic stellar mass) on various quantities and the method does not seem to work poorly in any particular regime and all trends are consistent with zero, see Appx.\,\ref{apx:err} and Fig.\,\ref{fig:err}.

In particular, no correlation was found between the mass residues and the galaxy distance. 
Regarding the angular size, no systematic trend is observed with radius (as represented by the \texttt{amaj} parameter) up to approximately 400\,arcsec. 
However, the eight largest galaxies (with radii between 518 and 803\,arcsec) exhibit a systematic stellar mass overestimation of 0.1\,--\,0.2\,dex. 
This is likely due to their extensive coverage across multiple Legacy Surveys tiles, which often results in poor-quality composite images. 
We would not advise to use our method on larger galaxies, however there are only a handful of galaxies larger than 803\,arcsec in the sky and they are often not even included in Legacy Surveys (e.g. M31 or Magellanic clouds). 

Similarly, no clear trend is observed with the galaxy inclination. 
Even with early-type galaxies excluded from our sample, we found no correlation between mass residues and axis ratios. 
However, a small subset of thin, edge-on galaxies with relatively low surface brightness tends to yield poorer mass estimates using our method (see Fig.\,\ref{fig:15w} for some examples). 

We examined various visual features of the galaxies in the sample and their correlations with the difference between S$^4$G masses and the mass estimates by Eq.\,(\ref{eq:f}). 
We specifically inspected galaxies with prominent dust lanes and found that their RMS is comparable to that of the general sample. 
Galaxies with strong tidal distortions tend to have slightly less accurate mass estimates, though they are not systematically disfavored by the method. 

Another subgroup of outliers consists of late-type galaxies with patchy star formation and nearby dwarf irregulars (see Fig.\,\ref{fig:15w} for some examples).
In these cases, substructures are often heavily masked by SEP as background or foreground sources, which likely leads to GALFIT models that do not accurately represent the galaxy total light distribution.

Interestingly enough, the faint thin edge-on galaxies and the late-type galaxies with patchy star formation (though not the nearby dwarf irregulars) also exhibit the highest uncertainties in Spitzer photometry, with errors exceeding 0.03\,mag.
This raises the question of whether the observed discrepancies originate from inaccurate Spitzer photometry or if the same factors contributing to the high uncertainties in Spitzer data also affect GALFIT photometry.

%%%%%%%%%%%%%%%%%%%%%%%%%%%%%%%%%%%%%%%%%%%%
%%%%%%%%%%%%%%%%%%%%%%%%%%%%%%%%%%%%%%%%%%%%
\section{Conclusions}

S$^4$G, as a deep mid-infrared survey, provides highly accurate stellar mass measurements. 
With the sample of 1732 S$^4$G galaxies, we calibrated relations that enable the estimation of stellar masses based on $g$, $r$, and $z$ band images from DR10 of the DESI Legacy Imaging Surveys. 
Integrated magnitudes from GALFIT 2-D S\'ersic model are recalculated to the extinction-corrected absolute magnitudes.
The most effective formula, Eq.\,(\ref{eq:f}), relies solely on absolute magnitudes in the $g$ and $r$ bands, reproducing the S$^4$G stellar masses with an RMS scatter of 25\%. 
It is expressed as: 
log($M_*$[M$_{\sun}$]) = $0.673M_g - 1.108M_r + 0.996$. 

Incorporating various combinations of magnitudes from all three bands, as well as structural parameters (S\'ersic index and axis ratio), failed to reduce the scatter below 24\%, see Appx.\,\ref{apx:forms}.
With slightly lower accuracy (a scatter of 29\%), Eq.\,(\ref{eq:sga}) can be used with \texttt{[G,R,Z]\_MAG\_SB[26]} magnitudes from the Siena Galaxy Atlas 2020 (SGA-2020), which are, however, available only for a portion of DR9 galaxies. 

The tables published in \citet{bell03} have been widely used for photometric estimates of stellar mass-to-light ratios in SDSS data.
However, when applied to the Legacy Surveys $g$ and $r$ images, their relation produces stellar masses with a scatter as high as 40\%.

Among the three bands employed in this work, the $z$-band wavelengths are the closest to Spitzer filters that were used in S$^4$G.
Despite that, the parameters extracted from the $z$-band images consistently brought the smallest improvements in reproducing the S$^4$G mass measurements. 
That holds for all GALFIT parameters (magnitudes, axis ratios, and S\'ersic indexes) as well as for SGA-2020 magnitudes.
The most helpful proved to be the $r$-band measurements. 

Systematic uncertainties arising from our choice of the reference sample are estimated to be at least 0.1\,dex. 
While our default calibration is based on \citet{eskew2012}, we also provide alternative fitting formulas based on stellar mass calibrations from \citet{meidt14} and \citet{querejeta15}, derived for both Legacy Surveys images as well as SGA-2020 magnitudes. 

We provide a Python-based script,  \texttt{photomass\_ls.py}\footnote{\url{https://github.com/PidiGalaxies/photomass}}, to automatically process the galaxies for the stellar mass estimates. 
The script requires the galaxy name and a radius (in arcseconds) as input parameters, representing the approximate angular size of the galaxy.
Optionally, the user can supply the galaxy distance (in Mpc) as a third input parameter, which can be accompanied with custom coordinates of the galaxy.
The script downloads images of the specified galaxy from the Legacy Surveys database, creates image masks, generates GALFIT input files with well-assessed initial values, performs the GALFIT photometry, and calculates the stellar mass estimate. 
If the galaxy distance is provided, the mass estimate uses this value, otherwise, it relies on NED redshifts. 
Additionally, an optional $K$ correction can be applied for redshifts up to 0.5.

Summary of limitations of our method:\\
(i) The Legacy Surveys images should be always checked for incomplete data or corrupted images (most often by a nearby bright star). For these purposes, script automatically downloads JPEG images.\\
(ii) Our method is verified to apparent total magnitudes up to 8.1 in the $g$ band.\\
(iii) We advise caution for the stellar mass estimates of galaxies larger than about 500\,arcsec.\\
(iv) Type of galaxies that often yield poor estimates:
\begin{itemize}[nosep]
  \item close overlaps with similar or bigger galaxies
  \item thin, edge-on galaxies with relatively low surface brightness
  \item late-type galaxies with patchy star formation and nearby dwarf irregulars
\end{itemize}
(v) Galaxies with prominent dust lanes and strong tidal distortions are not systematically disfavored by the method, although the latter tend to have slightly less accurate mass estimates.  
(iv) K-correction calculator implemented in the script works for redshift up to 0.5.

This work offers a robust approach to galaxy photometry and stellar mass estimation, providing valuable tools for the analysis of large galaxy samples.

%%%%%%%%%%%%%%%%%%%%%%%%%%%%%%%%%%%%%%%%%%%%
%%%%%%%%%%%%%%%%%%%%%%%%%%%%%%%%%%%%%%%%%%%%
\section{Data availability}
The stellar mass estimates, together with all basic and fitted parameters, are available in electronic form at Zenodo\footnote{\url{https://doi.org/10.5281/zenodo.14253992}}, and and at the CDS via anonymous FTP at cdsarc.u-strasbg.fr (130.79.128.5) or via the web interface\footnote{\url{http://cdsweb.u-strasbg.fr/cgi-bin/qcat?J/A+A/}}.

%%%%%%%%%%%%%%%%%%%%%%%%%%%%%%%%%%%%%%%%%%%%
%%%%%%%%%%%%%%%%%%%%%%%%%%%%%%%%%%%%%%%%%%%%
\begin{acknowledgements}
We would like to sincerely thank the anonymous referee for their thoughtful and constructive comments. Their careful reading, detailed critique, and concrete suggestions -- especially regarding the inclusion of alternative S$^4$G calibrations -- have significantly improved both the clarity and robustness of our work.
  	 \\ % MSCA
This project has received funding from the European Union's Horizon Europe Research and Innovation programme under the Marie Skłodowska-Curie grant agreement No. 101067618, GalaxyMergers.
  	\\ %MB
  	This work was supported by the Astronomical Station Vidojevica and by the Ministry of Science, Technological Development and Innovation of the Republic of Serbia through contract no. 451-03-66/2024-03/200002 made with the Astronomical Observatory of Belgrade.
  	\\ %NED
  	This research has made use of the NASA/IPAC Extragalactic Database, which is funded by the National Aeronautics and Space Administration and operated by the California Institute of Technology.
  	\\ %LS http://legacysurvey.org/acknowledgment/
The Legacy Surveys consist of three individual and complementary projects: the Dark Energy Camera Legacy Survey (DECaLS; Proposal ID \#2014B-0404; PIs: David Schlegel and Arjun Dey), the Beijing-Arizona Sky Survey (BASS; NOAO Prop. ID \#2015A-0801; PIs: Zhou Xu and Xiaohui Fan), and the Mayall z-band Legacy Survey (MzLS; Prop. ID \#2016A-0453; PI: Arjun Dey). DECaLS, BASS and MzLS together include data obtained, respectively, at the Blanco telescope, Cerro Tololo Inter-American Observatory, NSF’s NOIRLab; the Bok telescope, Steward Observatory, University of Arizona; and the Mayall telescope, Kitt Peak National Observatory, NOIRLab. Pipeline processing and analyses of the data were supported by NOIRLab and the Lawrence Berkeley National Laboratory (LBNL). The Legacy Surveys project is honored to be permitted to conduct astronomical research on Iolkam Du’ag (Kitt Peak), a mountain with particular significance to the Tohono O’odham Nation.
NOIRLab is operated by the Association of Universities for Research in Astronomy (AURA) under a cooperative agreement with the National Science Foundation. LBNL is managed by the Regents of the University of California under contract to the U.S. Department of Energy.
This project used data obtained with the Dark Energy Camera (DECam), which was constructed by the Dark Energy Survey (DES) collaboration. Funding for the DES Projects has been provided by the U.S. Department of Energy, the U.S. National Science Foundation, the Ministry of Science and Education of Spain, the Science and Technology Facilities Council of the United Kingdom, the Higher Education Funding Council for England, the National Center for Supercomputing Applications at the University of Illinois at Urbana-Champaign, the Kavli Institute of Cosmological Physics at the University of Chicago, Center for Cosmology and Astro-Particle Physics at the Ohio State University, the Mitchell Institute for Fundamental Physics and Astronomy at Texas A\&M University, Financiadora de Estudos e Projetos, Fundacao Carlos Chagas Filho de Amparo, Financiadora de Estudos e Projetos, Fundacao Carlos Chagas Filho de Amparo a Pesquisa do Estado do Rio de Janeiro, Conselho Nacional de Desenvolvimento Cientifico e Tecnologico and the Ministerio da Ciencia, Tecnologia e Inovacao, the Deutsche Forschungsgemeinschaft and the Collaborating Institutions in the Dark Energy Survey. The Collaborating Institutions are Argonne National Laboratory, the University of California at Santa Cruz, the University of Cambridge, Centro de Investigaciones Energeticas, Medioambientales y Tecnologicas-Madrid, the University of Chicago, University College London, the DES-Brazil Consortium, the University of Edinburgh, the Eidgenossische Technische Hochschule (ETH) Zurich, Fermi National Accelerator Laboratory, the University of Illinois at Urbana-Champaign, the Institut de Ciencies de l’Espai (IEEC/CSIC), the Institut de Fisica d’Altes Energies, Lawrence Berkeley National Laboratory, the Ludwig Maximilians Universitat Munchen and the associated Excellence Cluster Universe, the University of Michigan, NSF’s NOIRLab, the University of Nottingham, the Ohio State University, the University of Pennsylvania, the University of Portsmouth, SLAC National Accelerator Laboratory, Stanford University, the University of Sussex, and Texas A\&M University.
BASS is a key project of the Telescope Access Program (TAP), which has been funded by the National Astronomical Observatories of China, the Chinese Academy of Sciences (the Strategic Priority Research Program “The Emergence of Cosmological Structures” Grant \# XDB09000000), and the Special Fund for Astronomy from the Ministry of Finance. The BASS is also supported by the External Cooperation Program of Chinese Academy of Sciences (Grant \# 114A11KYSB20160057), and Chinese National Natural Science Foundation (Grant \# 12120101003, \# 11433005).
The Legacy Survey team makes use of data products from the Near-Earth Object Wide-field Infrared Survey Explorer (NEOWISE), which is a project of the Jet Propulsion Laboratory/California Institute of Technology. NEOWISE is funded by the National Aeronautics and Space Administration.
The Legacy Surveys imaging of the DESI footprint is supported by the Director, Office of Science, Office of High Energy Physics of the U.S. Department of Energy under Contract No. DE-AC02-05CH1123, by the National Energy Research Scientific Computing Center, a DOE Office of Science User Facility under the same contract; and by the U.S. National Science Foundation, Division of Astronomical Sciences under Contract No. AST-0950945 to NOAO.
  	\\ %SGA2020
The Siena Galaxy Atlas was made possible by funding support from the U.S. Department of Energy, Office of Science, Office of High Energy Physics under Award Number DE-SC0020086 and from the National Science Foundation under grant AST-1616414.
	\\ %k-correction calculator
This research made use of the ``K-corrections calculator'' service (kcor.sai.msu.ru).
	\\ %HyperLeda
We acknowledge the usage of the HyperLeda database (leda.univ-lyon1.fr).

\end{acknowledgements}

%%%%%%%%%%%%%%%%%%%% REFERENCES %%%%%%%%%%%%%%%%%
% join the .bib files when you upload your source files
\bibliographystyle{aa}
\bibliography{prolrot} % prolrot.bib

%%%%%%%%%%%%%%%%%%%%%%%%%%%%%%%%%%%%%%%%%%%%
%%%%%%%%%%%%%%%%%%%%%%%%%%%%%%%%%%%%%%%%%%%%
\begin{appendix}

%%%%%%%%%%%%%%%%%%%%%%%%%%
\section{Additional tested formulas} \label{apx:forms}

%------------------------------------------------------------------
\begin{table*} 
\caption{
Fitted formulas with additional parameters
} 
\label{tab:fits2} 
\centering
\begin{tabular}{lcccccccc}
\hline\hline
Formula & RMS & $a$ & $b$ & $c$ & $d$ & $e$ & $f$ & $g$ \\
\hline
$a M_g + b M_r + c q_g + d$ & 0.0957 & 0.660 & $-1.099$ & $-0.074$ & 0.989 & -- & -- & -- \\
$a M_g + b M_r + c q_r + d$ & 0.0959 & 0.656 & $-1.094$ & $-0.077$ & 0.986 & -- & -- & -- \\
$a M_g + b M_r + c q_z + d$ & 0.0961 & 0.654 & $-1.093$ & $-0.073$ & 0.982 & -- & -- & -- \\
$a M_g + b M_r + c q_g + d q_r + e$ & 0.0958 & 0.654 & $-1.092$ & 0.237 & $-0.316$ & 0.992 & -- & -- \\
$a M_g + b M_r + c M_z + d q_g + e$ & 0.0948 & 0.645 & $-1.058$ & $-0.023$ & $-0.067$ & 1.033 & -- & -- \\
$a M_g + b M_r + c M_z + d q_g + e q_r + f q_z + g$ & 0.0947 & 0.644 & $-1.057$ & $-0.022$ & 0.218 & $-0.306$ & 0.020 & 1.038 \\
\hline
$a M_g + b M_r + c n_g + d$ & 0.0967 & 0.703 & $-1.140$ & $-0.020$ & 0.975 & -- & -- & -- \\
$a M_g + b M_r + c n_r + d$ & 0.0958 & 0.729 & $-1.166$ & $-0.029$ & 0.964 & -- & -- & -- \\
$a M_g + b M_r + c n_z + d$ & 0.0971 & 0.680 & $-1.116$ & $-0.004$ & 0.992 & -- & -- & -- \\
$a M_g + b M_r + c n_g + d n_r + e$ & 0.0950 & 0.737 & $-1.174$ & 0.078 & $-0.103$ & 0.953 & -- & -- \\
$a M_g + b M_r + c M_z + d n_r + e$ & 0.0942 & 0.715 & $-1.120$ & $-0.029$ & $-0.030$ & 1.026 & -- & -- \\
$a M_g + b M_r + c M_z + d n_g + e n_r + f n_z + g$ & 0.0939 & 0.720 & $-1.128$ & $-0.027$ & 0.066 & $-0.094$ & 0.007 & 1.021 \\
\hline
$a M_g + b M_r + c q_g + d q_r + e n_g + f n_r + g$ & 0.0938 & 0.707 & $-1.146$ & 0.149 & $-0.208$ & 0.075 & $-0.091$ & 0.954 \\
\hline
\end{tabular}
\tablefoot{
Parameters and RMS values of different fitting formulas for calculating log($M_*$[M$_{\sun}$]) using galaxy total magnitudes, S\'ersic indexes, and axis ratios in different bands.
}
\end{table*}
%------------------------------------------------------------------

Here we present more complicated formulas that we tried when searching for the suitable relation to recover the stellar masses of galaxies in our sample. 
None of the formulas produced significant improvement. 
Compared to the 25.1\% scatter of the simple Eq.\,(\ref{eq:f}), the best result with more complex formulas lowered the RMS scatter only slightly, to 23.9\%.

Using only the absolute magnitudes, we constructed a second-degree polynomial equation with cross terms incorporating all colors
\begin{equation}
\begin{aligned}
\mathrm{log}(M_*[\mathrm{M}_{\sun}]) = & \ a M_z^2 + b M_r^2 + c M_g^2 + \\
& + d M_g M_r + e M_g M_z + f M_r M_z + \\
& + g M_g + h M_r + i M_z + j.
\label{eq:ply}
\end{aligned}
\end{equation}
The best fit with 21 outlying galaxies and parameter values [0.641, 0.324, -0.002, -0.975, -0.229, 0.245, 1.270, -1.883, 0.283, 1.851] have RMS of 0.0933\,dex (i.e. 24.0\% scatter).

In the next step we added the structural parameters provided by GALFIT -- the axis ratio, $q$ (marked “b/a” in GALFIT outputs), and the S\'ersic index, $n$, which is a good proxy for the galaxy morphological type.
We tested various combinations of these quantities, focusing on enhancing the most effective formula Eq.\,(\ref{eq:f}), identified in Sect.\,\ref{ssec:formula}.  
The improvement in RMS was consistently subtle across all cases, as shown in Tab.\,\ref{tab:fits2}.  
Within these modest changes, including the S\'ersic index or the axis ratio for a single band showed a slightly better preference for the axis ratio.  
However, formulas incorporating more color bands tended to favor the S\'ersic index. 
Once again, the $z$ band proved to be the least helpful for both additional parameters. 

Combining both -- the S\'ersic index and the axis ratio -- simultaneously did not yield significant improvements.  
Even extending the combination to include all three quantities across all three bands  
\begin{equation}
\begin{aligned}
\mathrm{log}(M_*[\mathrm{M}_{\sun}]) = & \ a M_g + b M_r + c M_z + \\
& + d q_g + e q_r + f q_z + \\
& + g n_g + h n_r + i n_z + j.
\label{eq:qn}
\end{aligned}
\end{equation}
with parameter values [0.697, -1.111, -0.022, 0.180, -0.257, 0.027, 0.064, -0.085, 0.006, 1.015] improved the RMS value only to 0.0932\,dex (i.e. 23.9\% scatter).

%%%%%%%%%%%%%%%%%%%%%%%%%%
\section{Residual dependences} \label{apx:err}

%------------------------------------------------------------------
\begin{figure*}
    \centering
    \includegraphics[width=1\linewidth]{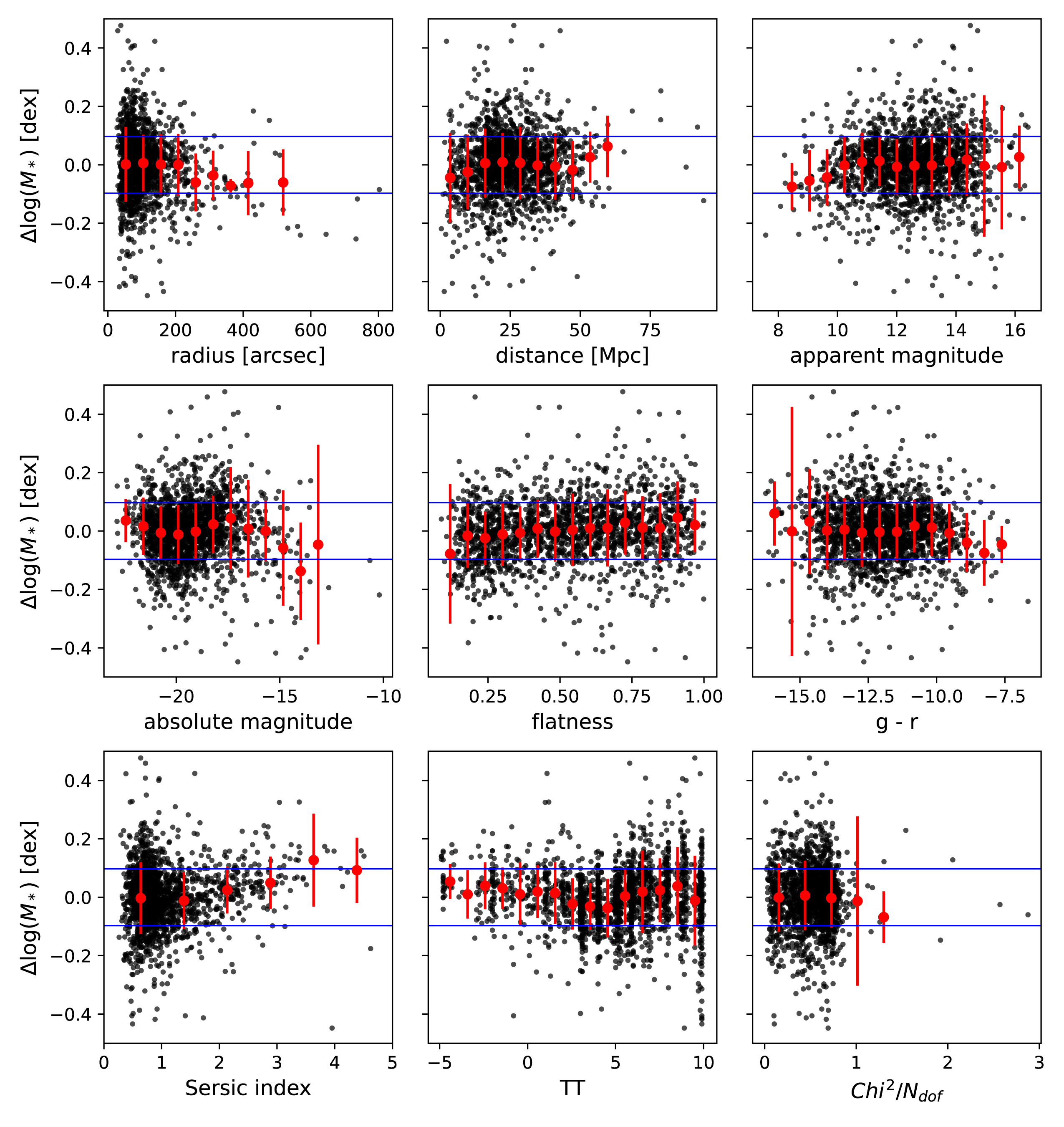}
    \caption{ 
Dependence of mass residues (i.e. estimated logarithmic stellar mass minus the reference logarithmic stellar mass) on different quantities, see Appx.\,\ref{apx:err}. 
Red dots with error bars show the median value in the given bin and its standard deviation. 
Blue horizontal lines denote the RMS value of $\pm 0.0972$\,dex of our fitted relation, Eq.\,(\ref{eq:f}).
}
    \label{fig:err}
\end{figure*}
%------------------------------------------------------------------

We checked the dependence of mass residues (i.e. estimated logarithmic stellar mass minus the reference logarithmic stellar mass) on various quantities: 
\begin{itemize}
  \item radius of the galaxies (the \texttt{amaj} parameter)
  \item galaxy distance
  \item apparent magnitude in $g$ band
  \item absolute magnitude in $g$ band
  \item flatness, i.e. the axis ratio provided by GALFIT
  \item  $g-r$ color
  \item S\'ersic index provided by GALFIT
  \item TT morphological type code provided by S$^4$G catalog
  \item GALFIT indicator of goodness of the fit $Chi^2/N_{\mathrm{DOF}}$
\end{itemize}
All trends are consistent with zero, see Fig.\,\ref{fig:err}. 
However, there are indications that very large galaxies (with radii exceeding 500\,arcsec) may have their stellar masses systematically underestimated (by about 0.2,dex), while galaxies with high S'ersic indices tend to show systematic overestimations.

%%%%%%%%%%%%%%%%%%%%%%%%%%
\section{Using the script} \label{apx:ex}

The \texttt{photomass\_ls.py} script is designed for use with Python\,3.
An example of running the script for the galaxy NGC\,474 with radius of 106.3\,arcsec and the distance of 30.88\,Mpc:
\begin{verbatim}
python3 photomass_ls.py NGC474 106.3 --dist 30.88
\end{verbatim}
The output:
\begin{verbatim}
Galaxy: NGC474
RA: 20.0279 Dec: 3.416
log10(M*[Sun]): E: 10.79 M: 10.84 Q: 10.88
Ext[mag]: g : 0.112 r : 0.075
Mag[mag]: g : -20.5 r : -21.29
Sersic index: g : 3.049 r : 3.872
R_e[px]: g : 44.73 r : 48.04
R_e[arcsec]: g : 23.44 r : 25.17
Axis ratio: g : 0.8185 r : 0.8188
Position angle[deg]: g : 21.18 r : 19.1
Distance[Mpc]: 30.88 (from input)
Plate scale[arcsec / px]: 0.524 0.524
Zero point[mag]: 20.99
Redshift: 0.007722
\end{verbatim}
To get help, type:
\begin{verbatim}
python3 photomass_ls.py -h
\end{verbatim}

\end{appendix}

%%%%%%%%%%%%%%%%%%%%%%%%%%%%%%%%%%%%%%%%%%%%
\label{LastPage} % to be inserted before \end{document}. Repairs a bug with multiple citations in aa.cls
\end{document}